\documentclass[twocolumn]{aa} 
\usepackage{graphicx}
\usepackage{txfonts}

\def\kms {\ifmmode{{\rm \ts km\ts s}^{-1}}\else{\ts km\ts s$^{-1}$}\fi}
\begin{document}
\title{On the nature of dust clouds in the region towards M\,81 and NGC\,3077}
\authorrunning{Andreas Heithausen}
\titlerunning{Dust clouds towards M\,81 and NGC\,3077}
\author{Andreas Heithausen}
\institute{I. Physikalisches Institut, Universit\"at K\"oln, Z\"ulpicher 
Str. 77, 50937 K\"oln, Germany, \email{aheithau@uni-koeln.de}   
\thanks{New address: Institut f\"ur Physik und ihre Didaktik,
  Universit\"at zu K\"oln, Gronewaldstr. 2, 50931 K\"oln, Germany}} 
\date{Received 9 Aug 2011; accepted 18 May 2012 }

\abstract
{}
{There is some controversy on the nature of dust clouds found in
  direction of the interacting galaxy triplett M\,81, M\,82, and
  NGC\,3077. Are they associated with the tidal arms seen in HI 
  around those galaxies or are they simply Galactic foreground clouds?} 
{Data from the SPIRE instrument onboard HERSCHEL\thanks{{\it
  Herschel} is an ESA space observatory with science instruments
  provided by European-led Principal Investigator consortia and with
  important participation from NASA.} and MIPS onboard of SPITZER 
are used to
  derive physical parameters for the dust clouds. These observions are
  compared to CO clouds previously mapped with the IRAM and the FCRAO
  radio telescopes.}
{SPIRE and MIPS maps show several dust clouds north of M\,81 and south
  of NGC\,3077. Modelling of the dust emission provides total hydrogen
  column densities between 1.5 and $5.0\cdot 10^{20}$\,cm$^{-2}$. Dust
  temperatures are between 13 to 17\,K. No significant difference in
  the dust emission can be found between individual clouds.  It is
  shown that CO line emission provides the best clues on the
  origin of those clouds. Most of the clouds seen towards M\,81 are
  associated with small-area molecular structures (SAMS), i.e. tiny CO clouds
  of Galactic origin. The clouds seen towards NGC\,3077 are partly
  associated with the tidal arms and are partly in the Galactic 
  foreground associated with SAMS.}
  {}
   
\maketitle


\section{Introduction}

Interpretation of astronomical observations is always hampered by the
lack of direct distance information. Whether or not objects on the
same line-of-sight are indeed physically related is not easy to
judge. Directly linked to this question is the problem of finding dust
in the intergalactic medium (Xilouris et al. \cite{xilouris:etal06};
Walter et al. \cite{walter:etal11}). One well studied region where
this problem becomes obvious is the region towards \object{M\,81}, 
M\,82, and \object{NGC\,3077}. It is one of the
closest sample of interacting galaxies. HI tidal arms are found to
connect these three galaxies (Yun et al. \cite{yun:etal94}).
Especially since the discovery of Arp's loop (Arp \cite{arp65})
 there has been some debate about the nature of interstellar clouds in that
 region: are they related
to tidal arms around the interacting galaxy triplett (Sun et
al. \cite{sun:etal05}; de Mello et al. \cite{demello:etal08}) or to
Galactic foreground cirrus (Sollima et al. \cite{sollima:etal10};
Davies et al.  \cite{davies:etal10})?

Sandage (\cite{sandage76}) presented deep
optical images that showed that the M\,81-M\,82-NGC\,3077 region is
seen through wide spread Galactic foreground cirrus clouds. De Vries
et al. (\cite{devries:etal87}) presented large-scale maps with an angular 
resolution of 9\,arcmin of atomic
hydrogen, carbon monoxide and dust infrared emission, which showed Galactic
cirrus emission towards the M\,81 triple with total hydrogen column
densities of about $1-2 \cdot 10^{20}$cm$^{-2}$. 
In most cases it is possible to distinguish between an extragalactic or
  Galactic origin of the dust and gas because the radial velocities are very
  different from each other.  In the case of the M\,81 region however the
  radial velocities of Galactic and extragalactic gas share (at least partly)
  the same LSR velocity range close to zero.

Significant substructure in the Galactic foreground gas was detected
by Heithausen (\cite{heithausen02}) who found several small-area
molecular structures (SAMS) towards the outer regions of M\,81 and
NGC\,3077. SAMS are tiny molecular clouds detected in a region where
the shielding of the interstellar radiation field is too low for them
to survive for a long time (Heithausen \cite{heithausen02}). The CO
clouds are less than one arcmin wide, corresponding to linear sizes
below 6000\,AU at an adopted distance of 100pc. High angular
resolution observations obtained with the Plateau de Bure
interferometer show substructure down to the resolution limit of
3$''$\ (approx. 300\,AU) with indication of further unresolved
structure (Heithausen \cite{heithausen04}).

So far, one such cloud has been detected towards the tidal arms around
NGC\,3077 and four in the region around M\,81 (Heithausen
\cite{heithausen06}). The agreement of the LSR velocities of SAMS with
that of the Galactic HI gas shows that they are of Galactic
origin. Furthermore their CO linewidths are too narrow for
extragalactic objects. At an adopted distance of about 100pc their
masses are about that of Jupiter.

\begin{figure*}
\includegraphics[angle=-90,width=16.5cm]{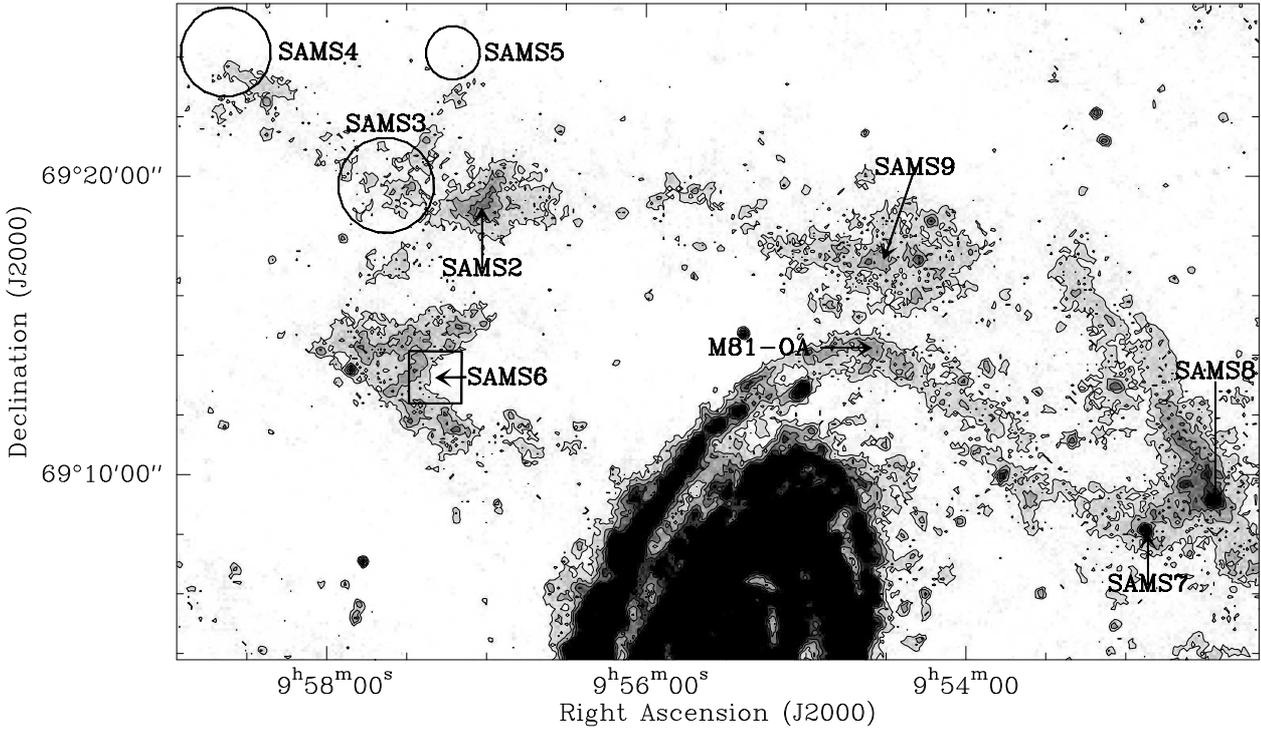}
\caption
{SPIRE 250 $\mu$m map of the area surrounding SAMS2. Contours are from
  0.02 to 0.1\,Jy\,beam$^{-1}$ every 0.02\,Jy\,beam$^{-1}$. The
  intense emission at the lower centre of the map originates from
  M\,81. The positions of the sources discussed in this paper are
  labeled. The weak clouds SAMS 3-5 are marked by circles with 
  the approximate size of the clouds. The square marks the area of SAMS6 over
  which spectra have been averaged to detect a weak CO line (s. Fig. \ref{sams6-fcrao}).}
\label{sams2-spire250}
\end{figure*}
\begin{figure*}
\includegraphics[angle=-90,width=16.5cm]{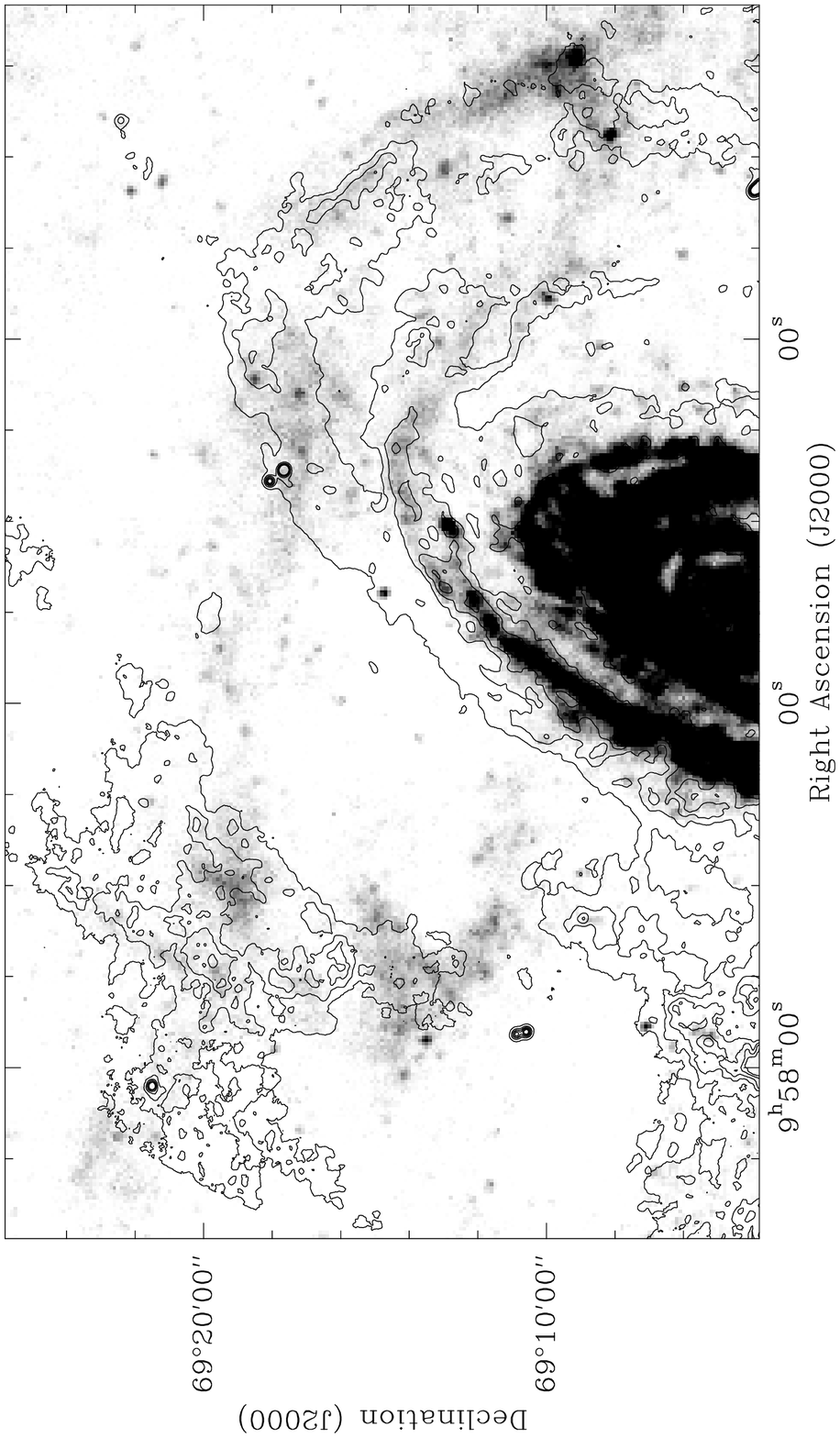}
\caption
{THINGS total HI map (contours, Walter et al.  \cite{walter:etal08}) of the
  area surrounding M\,81 overlayed on the SPIRE 250\,$\mu$m map
  (greyscale). The map has been integrated over a velocity range of about
  250\,km\,s$^{-1}$.
Contours are from 50 to 500\,Jy\,beam$^{-1}$\,m\,s$^{-1}$ 
  every 100\,Jy\,beam$^{-1}$\,m\,s$^{-1}$. }
\label{m81-hi}
\end{figure*}

The purpose of this paper is twofold: first independent estimates on
column densities and temperatures of SAMSs are obtained, second these
objects are used to distinguish between Galactic foreground clouds and
extragalactic tidal arm clouds. The outline of the paper is as
follows: after a brief description of the infrared data the resulting
spectral energy distributions (SEDs) are used to determine dust column densities
and temperatures and find new SAMS candidates; one of those could
directly be confirmed as molecular cirrus cloud with existing CO
data. Based on published CO data the dust clouds are classified as
Galactic or extragalactic. It is shown that the dust clouds towards
M\,81 are most likely Galactic foreground clouds, whereas the clouds
seen towards NGC\,3077 are partly associated with the tidal arms and
are partly in the Galactic foreground.

\section{Observations\label{observations}}
For the analysis in this paper I used photometric data obtained with the
  SPIRE (Griffin et al. \cite{griffin:etal10}) oboard of the  Herschel satellite and with MIPS
  (Rieke et al. \cite{rieke:etal04}) onboard of the Spitzer satellite.
  Calibrated data of the M\,81 region obtained with SPIRE  were retrieved
  from the Herschel archive (level 2 products).  A first analysis of the data
  is presented by Bendo et al. (\cite{bendo:etal10}) to study the dust content
  and temperature of the spiral galaxy. SPIRE data of NGC\,3077 were obtained
  from the open time key project KINGFISH (Walter et
  al. \cite{walter:etal11}).

The SPIRE data are given in Jy\,beam$^{-1}$; to convert the data to MJy
sr$^{-1}$\ beam areas of 447, 816, and 1711 arcsec$^{-1}$\ were assumed for
the 250, 350, and 500$\mu$m bands (Sibthorpe et al. \cite{sibthorpe:etal11}). 
The calibration uncertainties for the individual bands are 7\% (SPIRE
Observor's Manual \cite{spire2011}). 
At 250, 350 and 500$\mu$m SPIRE has a slightly elliptical beam of
 $18.7''\times17.5''$, 
$25.6''\times 24.2''$ and $38.2''\times 34.6''$, respectively (Sipthorpe et
 al., loc.cit.) for pixel scales of 6, 10, and 14 arcsec.

For the M\,81 region photometric data at 70 and 160 $\mu$m obtained with
Spitzer MIPS were retrieved from the SINGS archive (Kennicutt et
al. \cite{kennicutt:etal03}). Corresponding data at 160 $\mu$m for the region
around NGC\,3077 were taken directly from the Spitzer archive. For this region
three data sets exist in the archive which cover the clouds of interest
  completely, though part of the more extensive diffuse emission around the
  galaxy might be
  missing. Because the signal-to-noise level of a single set is already high
enough (larger than 10 for the weakest cloud) I only used one data set (AOR
17597952), after checking that the data set is comparable with the other sets
(AORs 17598208 and 17597696). The data sets are calibrated in MJy/sr.  The
FWHM of the point-spread function (PSF) is 18$''$, and 40$''$ at 70 $\mu$m and
160 $\mu$m, resp. (Engelbracht et al. \cite{engelbracht:etal04}). The accuracy
of the photometric calibration of the two bands was estimated to be 5\%\ at
70\,$\mu$m (Gordon et al. \cite{gordon:etal07}) and 12\% at 160\,$\mu$m
(Stansberry et al. \cite{stansberry:etal07}).

Additionally to the photometric observations, 
the IRAM 30\,m telescope has been used to search for extensive $^{12}$CO
$(1\to0)$ emission around SAMS1 in June 2001. Observations were done with a
wobbling subreflector with an off-position separated by $200''$ in azimuth
from the on-position.  An autocorrelator spectrometer was used with a velocity
resolution of 0.2 km s$^{-1}$.  The angular resolution of the telesccope at
115 GHz was 22$''$. The data were processed using the standard data
  reduction software for radioastronomical spectra 
GILDAS\footnote{see http://www.iram.fr/IRAMFR/GILDAS/}. Only linear
  baselines were removed from the spectra.

\section{Results\label{results}}

\subsection{The region towards M\,81}

The SPIRE 250$\mu$m map of the M\,81 region is presented in
Fig. \ref{sams2-spire250}. Dust is concentrated in several small
filaments. Six clouds are identified outside the main body of
M\,81. Positions of the local maxima are listed in Tab.
\ref{spire-clouds}. 

For comparison the  distribution of the total HI gas
of this region taken from the THINGS project (Walter et
al. \cite{walter:etal08}) is displayed in Fig. \ref{m81-hi}. 
The velocity  range of the naturally weighted maps covers about 250 
km\,$s^{-1}$. Note that 
because the interferometer data are not corrected for missing zero spacing 
extensive HI emission has been filtered out. Due to the small size of the 
dust clouds this is however not critical for the analysis. The distribution of
HI
associated with the spiral arms of M\,81 shows a close similarity to
the dust. The HI clouds associated with the tidal arms of M\,81 show however
no obvious morphological similarity to the dust clouds. 
Davies et al. (\cite{davies:etal10}) already noted that a much greater
  similarity can be found when comparing the infrared data with a single
  velocity channel of the HI data at velocities close to zero, i.e. that which
  correspond to  Galactic velocities.

\begin{figure}
\includegraphics[angle=-90,width=9cm]{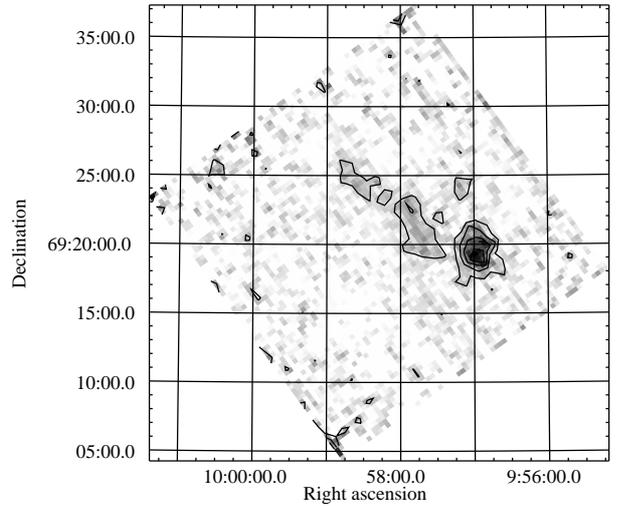}
\caption
{Integrated CO map obtained with the FCRAO 14\,m radiotelescope
(Heithausen \cite{heithausen06}). Contours are every 0.04 K km s$^{-1}$\ 
starting at 0.04 K km s$^{-1}$. }
\label{sams-fcrao}
\end{figure}

A morphological similarity can be found when one compares the dust with
the CO emission from small-area molecular structures, too. \object{SAMS2}
(Heithausen \cite{heithausen02}) can clearly be identified in all
SPIRE bands (compare Figs. \ref{sams2-spire250} and \ref{sams-fcrao})
and in the MIPS 160$\mu$m band. The other molecular structures
in its vicinity found in CO observations with the FCRAO 14m telescope
(Heithausen \cite{heithausen06}) are detected in the 250$\mu$m map, too.
 Note that the CO data have a coarser angular 
resolution of only $45''$ compared to the SPIRE 250$\mu$m data and 
are very weak ($T_A^*<0.1$\,K).

\begin{figure}
\includegraphics[angle=-90,width=8.5cm]{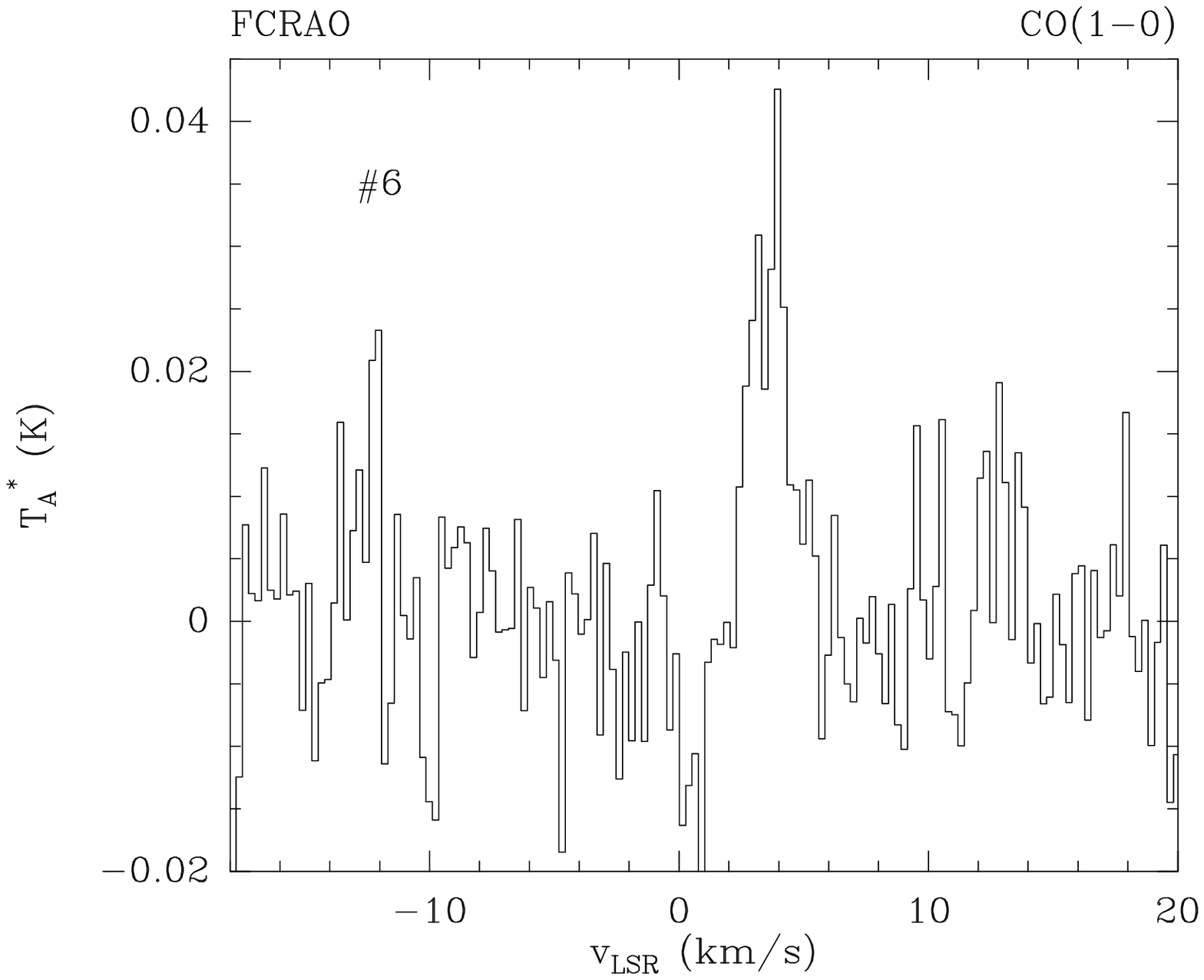}
\caption
{CO $1\to0)$ spectrum of the newly detected SAMS6 centered on
 $(\alpha,\delta)=(09^{\rm h}57^{\rm m}18^{\rm s}\!.5,
+69^\circ13'22''\!.4)$ }
\label{sams6-fcrao}
\end{figure}

There is one further dust cloud in the FCRAO field (compare Fig. 
\ref{sams2-spire250} and \ref{sams-fcrao}) previously
undetected in CO (labelled SAMS6 in Fig. \ref{sams2-spire250}). 
To see whether molecular gas is associated with
that cloud the spectra obtained with the FCRAO 14m telescope 
(Heithausen \cite{heithausen06}) were averaged over areas of 60$''$ by
60$''$. In those spectra four neighboring positions with marginal CO
detections can be found centered on the Galactic coordinates
$(l,b)=(141^\circ\!.802,40^\circ\!.936)$\ which corresponds to
$(\alpha,\delta)=(09^{\rm h}57^{\rm m}18^{\rm s}\!.5,
+69^\circ13'22''\!.4)$. The average of those positions is shown in
Fig. \ref{sams6-fcrao}. Here a clear CO line has been detected with
the following values obtained from a Gaussian fit to the data:
amplitude $T_{\rm A}^* = 0.032 \pm 0.008$\,K, center velocity $v_{\rm
  LSR}=3.6 \pm 0.1$\,km\,s$^{-1}$, and line width $\Delta v = 1.8
\pm0.3$\,km\,s$^{-1}$. These values are similar to those of previous 
detected SAMS. 

For the further quantitative analysis (Sec. \ref{section-seds}) six
apparent cirrus clouds are identified, three of which show CO emission
of clear Galactic origin (SAMS2, 4, and 6) while for the other three
such observations are not yet available (labelled SAMS7, 8, and 9 in
Tab. \ref{spire-clouds} and Fig. \ref{sams2-spire250}). 
Although SAMS3 and SAMS5 show CO emission, they are not considered further
because they are hard to identify in the SPIRE 350 and 500$\mu$m bands.
For comparison one cloud in an outer spiral arm of M\,81 (labelled
M\,81-OA is also listed in Tab. \ref{spire-clouds}).

\subsection{The region towards NGC\,3077}

\begin{figure}
\includegraphics[angle=-90,width=8.5cm]{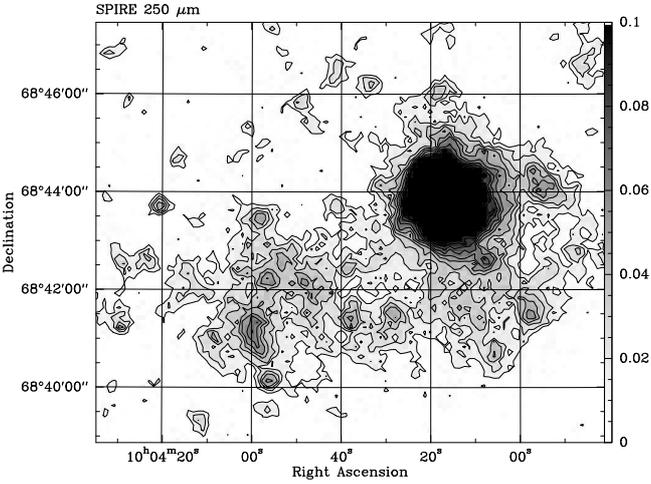}
\caption
{Spire 250 $\mu$m map of the region towards NGC\,3077 as observed by
  Walter et al. (\cite{walter:etal11}). Contours are from 0.01 to 0.08
  Jy\,beam$^{-1}$ every 0.01 Jy\,beam$^{-1}$. The intense emission in
  the right half of the map originates from NGC\,3077.}
\label{spire-ngc3077}
\end{figure}

NGC\,3077 is the smallest member of the interacting M\,81 galaxy
triple.  Intense HI tidal arms have been found surrounding this galaxy
(Yun et al.  \cite{yun:etal94}). Towards these arms Walter \&
Heithausen (\cite{walter:heithausen99}) have detected a giant
molecular complex.  Walter et al. (\cite{walter:etal11}) have used the
SPIRE instrument to search for the thermal emission of dust asociated
with the tidal arms.  A map at a wavelength of 250$\mu$m is presented
in Fig. \ref{spire-ngc3077}. Based on a positional coincidence Walter
et al. attribute all of the emission they found in that region to the
tidal arms around NGC\,3077. In this section it is shown that at least
some part is most likely associated with Galactic foreground clouds.

\begin{figure*}  
\sidecaption
\includegraphics[angle=0,width=13cm]{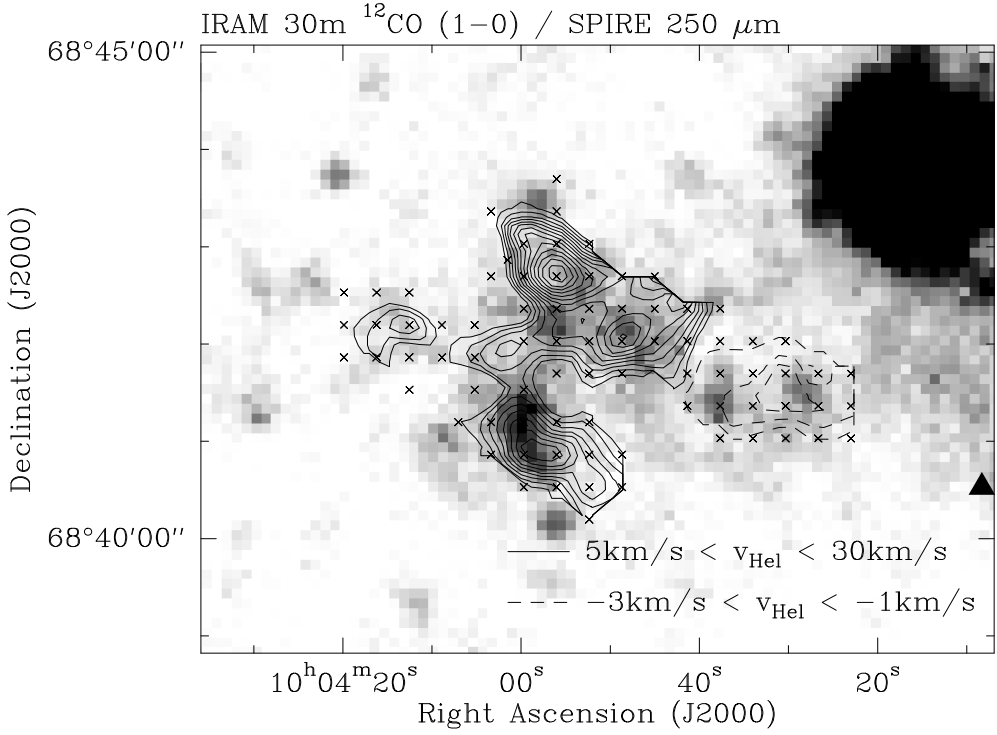}
\caption
{Spire 250 $\mu$m map of the region towards NGC\,3077 overlaid with
  contours from observations of the CO ($1\to0$) line obtained with
  the IRAM 30\,m telescope (Heithausen \cite{heithausen02}; Walter \&
  Heithausen \cite{walter:heithausen99}; Heithausen \& Walter
  (\cite{heithausen:walter00}). Observed CO positions are marked by
  Xs. Contours are from 0.1\,K\,km\,s$^{-1}$ every
  0.1\,K\,km\,s$^{-1}$. Solid contours show emisson from the tidal
  arm cloud associated with NGC\,3077 (Walter \& Heithausen
  \cite{walter:heithausen99}; Heithausen \& Walter
  \cite{heithausen:walter00}), whereas dashed contours show emission
  from the Galactic small area molecular structure SAMS1 (Heithausen
  \cite{heithausen02}). There is further CO emission towards the position
  $(\alpha,\delta)=(10^{\rm h}03^{\rm m}08^{\rm s}\!.4, +68^\circ40'31''\!.8)$, 
  which is marked by the black triangle. }
\label{phoenix-sams1}
\end{figure*}

In the region towards NGC\,3077 we are faced with two molecular clouds,
one of clear extragalactic and one of clear Galactic origin. The situation
is displayed in Fig. \ref{phoenix-sams1}. The distinction
can be made based on the CO linewidth and LSR velocity. The extragalactic cloud
complex (Walter \& Heithausen \cite{walter:heithausen99}, Heithausen
\& Walter \cite{heithausen:walter00}) shows CO linewidths of several
km\,s$^{-1}$\ and velocities slightly outside that of the local
Galactic gas (solid contours). Next to this complex is the small-area
molecular structure \object{SAMS1} (dashed contours, Heithausen 
\cite{heithausen02}) with a linewidth of below 1\,km\,s$^{-1}$\ and 
LSR velocities in agreement with that of local Galactic HI gas.

Beyond the CO map of SAMS1 presented in Fig. \ref{phoenix-sams1} 
further CO $(1\to0)$ emission has been detected 
with the IRAM 30m telescope
towards the position $(\alpha,\delta)=(10^{\rm h}03^{\rm
  m}08^{\rm s}\!.4, +68^\circ40'31''\!.8)$.  A Gaussian fit to the
data gives a amplitude $T_{\rm A}^* = 0.055 \pm 0.004$\,K, a center
velocity $v_{\rm LSR}=3.47 \pm 0.14$\,km\,s$^{-1}$, and a line width
$\Delta v = 1.6 \pm0.3$\,km\,s$^{-1}$. (The LSR velocity corresponds to a 
heliocentric velocities of $v_{\rm Hel}=-2.53$\,km\,s$^{-1}$.) These values 
are similar to that of SAMS1, i.e. the molecular gas is of Galactic origin.

Both the Galactic and the extragalactic clouds are associated with
emission in the 250\,$\mu$m band (s. Fig. \ref{phoenix-sams1}) as well
as with emission in the 350 and 500\,$\mu$m band. 
For further quantitative analysis three
extragalactic regions (NGC\,3077-TA1 to TA3) and one Galactic cloud
(SAMS1) are identified (see Tab. \ref{spire-clouds}). These positions are
associated with the most intense regions in this SPIRE 250\,$\mu$m map
outside the main body of NGC\,3077.

\begin{table*}
\caption{Parameters of the dust clouds derived from SPIRE and MIPS data}
\label{spire-clouds}
\centering
\begin{tabular}{l c c c c c c c c c}
\hline\hline
 Source & RA       & Dec   &  $I_{250\mu \rm m}$ &
 $I_{350\mu \rm m}$ &  $I_{500\mu \rm m}$  &  $I_{70\mu \rm m}$ &  $I_{160\mu
   \rm m}$ &$T_{\rm Dust}$ & $N_{\rm H}$ \\
        & J2000  &J2000&MJy sr$^{-1}$&MJy sr$^{-1}$&MJy
 sr$^{-1}$&MJy sr$^{-1}$ &MJy sr$^{-1}$&K& $10^{20}$\,cm$^{-2}$ \\
              &           &            &SPIRE&SPIRE&SPIRE& MIPS &MIPS \\
\hline
SAMS1         &10:03:28.4 & 68:41:29.2 & 3.9 & 2.1 & 1.0 & ...      & 4.5 & 14.5 & 1.8 \\
NGC\,3077-TA1 &10:04:00.6 & 68:41:09.9 & 5.7 & 3.6 & 2.0 & ...     & 4.9 & 13   & 5.4 \\
NGC\,3077-TA2 &10:03:55.4 & 68:42:39.4 & 4.2 & 2.7 & 1.5 & ...      & 5.1 & 13   & 3.5 \\
NGC\,3077-TA3 &10:03:48.0 & 68:42:01.4 & 4.7 & 2.8 & 1.6 & ...      & 5.1 & 13   & 3.6\\
SAMS2         & 9:57:01.0 & 69:19:05.4 & 7.6 & 3.5 & 1.5 & $\le0.5$ & 7.4 & 16   & 3.3 \\
SAMS3         & 9:57:36.8 & 69:19:48.7 & 3.6 & ... & ... & ...      & ... & ...  &...  \\
SAMS4         & 9:58:36.5 & 69:24:16.8 & 4.6 & 2.3 & 0.9 & ...      & ... & 13.5 & 3.7 \\
SAMS5         & 9:57:11.9 & 69:24:15.3 & 2.7 & ... & ... & ...      & ... & ...  &...  \\
SAMS6         & 9:57:18.5 & 69:13:22.4 & 5.7 & 3.0 & 1.5 & $\le0.5$ & 6.7 & 15.5   & 3.1 \\
SAMS7         & 9:52:53.1 & 69:08:09.6 & 8.6 & 3.8 & 1.5 & 2.9$^1$  & 8.0 & 16.5   & 3.1 \\
SAMS8         & 9:52:28.0 & 69:09:06.8 & 11.2& 5.2 & 1.9 & $\le0.5$ & 10.9& 14.5 & 7.2 \\
SAMS9         & 9:54:31.4 & 69:17:19.5 & 6.3 & 3.4 & 1.6 & $\le0.5$ & 5.4 & 13   & 6.9 \\
M\,81-OA      & 9:54:35.5 & 69:14:22.0 & 6.5 & 3.6 & 1.3 & $\le0.5$ & 5.5 & 14   & 5.1 \\
\hline
\end{tabular}
\tablefoot{1: possible contributions from a point source}
\end{table*}

\subsection{SEDs \label{section-seds}}
To obtain dust column densities and temperatures the IR maps were smoothed to
the same angular resolution and intensities averaged
over square areas with 40$''$ width were obtained.  Background
correction were obtained in fields close to the observed regions which
show no obvious emission. Derived intensities for the individual spectral bands
are listed in Tab. \ref{spire-clouds}. 
The uncertainties for the values
are estimated from the noise level (1$\sigma$) in regions, which are
apparently free of emission.
For the M\,81 region these values are found to be at 0.60, 0.38 and 0.21
MJy sr$^{-1}$ for the Herschel 250, 350, and 500$\mu$m bands and
0.5\,MJy\,sr$^{-1}$ for both the MIPS 70 and the 160$\mu$m bands.
For the NGC\,3077 region the correspoinding values are 0.55, 0.27 and 0.16 MJy
sr$^{-1}$ for the Herschel 250, 350, and
500$\mu$m bands and 0.5 MJy sr$^{-1}$  for the MIPS 160$\mu$m band,
respectively.

\begin{figure*}
\includegraphics[angle=-90,width=18cm]{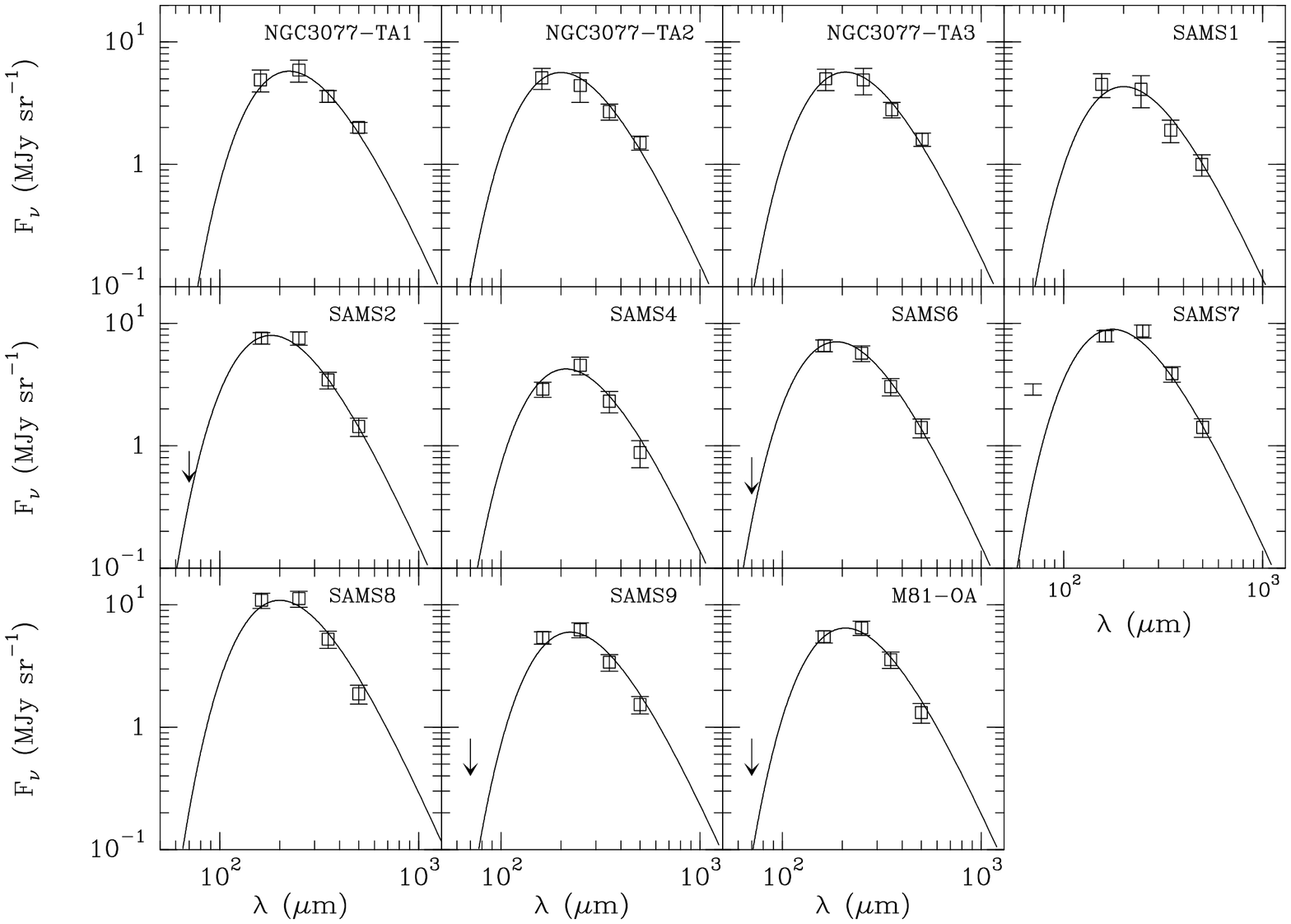}
\caption
{Dust spectra of the observed sources obtained from the SPIRE and 
  MIPS bands.
Each point gives the mean intensity of a field covering an area of 
$40'' \times 40''$. Solid lines are fits to the data. 
For the fits $\beta=2$ has been adopted. The 160\,$\mu$m intensity of SAMS7
shows contributions from a point source.}
\label{seds}
\end{figure*}

A blackbody fit is used to derive dust temperatures and hydrogen column 
densities of the observed regions. Assuming optically thin emission 
the observed intensity $I_\nu$ can be modelled as 
$$ I_\nu= B_\nu \cdot m_{\rm H} \cdot\mu \cdot N_{\rm H} \cdot\kappa_\nu$$

where $B_\nu$ is the Planck function and $N_{\rm H}$ is the total hydrogen
column density (cf. Ward-Thompson et al. \cite{ward-thompson:etal10}).
  $m_{\rm H} \cdot\mu$ is the mean mass of a particle in the cloud with
  $m_{\rm H}$ being the mass of an hydrogen atom; $\mu$ was adopted to be 2.3
  (cf. Ade et al. \cite{ade:etal11}).  Following Beckwith et
al. (\cite{beckwith:etal90}) the dust mass opacity $\kappa_\nu$ is
parameterized as
$$ \kappa_\nu=0.1 \rm{cm^2 g^{-1} \cdot \left(\nu \over {1000GHz}
  \right)^\beta}$$ For the fit a dust opacity index $\beta=2$ has been
adopted,  which is consistent with the value used by Andr\'e et
  al. (\cite{andre:etal93}) for molecular cores.  Fits to the data are
displayed in Fig. \ref{seds}; derived temperatures and total hydrogen column
densities are listed in Tab. \ref{spire-clouds}.
 
The derived dust temperatures are within the range of 13 to 17\,K 
with an uncertainty of $\Delta T = \pm2$\,K for individual values.
There is no significant difference between the dust temperature of the
outer arm of M\,81, the tidal arm features close to NGC\,3077, and the
Galactic foreground clouds. The values are similar to those derived for
typical cirrus clouds as e.g. the Polaris Flare where the average dust
temperature was found to be $T_D=14.5\pm1.6$\,K (Miville-Desch\^enes
et al. \cite{miville:etal10}). These values agree also well with
those estimated from exciation conditions of the CO line between 7
to 20\,K (Heithausen \cite{heithausen04}), 
although gas and dust are not necessarily coupled.

The total hydrogen column densities estimated from the SED fit are
within 1.8 and $7.2 \cdot 10^{20}$\,cm$^{-2}$. The uncertainty is 
estimated to be $\pm 0.5 \cdot 10^{20}$\,cm$^{-2}$. Again no significant
difference between extragalactic and Galactic clouds is apparent. For
SAMS1 Heithausen (\cite{heithausen02}) found an HI column density of
$N$(HI)=$1.3\cdot10^{20}$\,cm$^{-2}$ and an H$_2$ column densities of
$N$(H$_2)=0.7\cdot 10^{19}$\,cm$^{-2}$ and for SAMS2
$N$(HI)=$2.1\cdot10^{20}$\,cm$^{-2}$ and $N$(H$_2)=3.7\cdot
10^{19}$\,cm$^{-2}$. The sums of $N(\rm H)= N(\rm HI)+2\cdot
N(\rm H_2)$ agree well with the values derived here from the IR 
emssion for both clouds.

\section{Discussion}

\subsection{SAMS}

Previous estimates on the kinetic temperatures and masses of SAMSs
(Heithausen \cite{heithausen02, heithausen06}) were made using HI and
CO data assuming that the physical parameters are similar to those of
Galactic cirrus clouds. While the determination of the atomic hydrogen
column density is straightforward from its 21cm line, the determination
of the molecular hydrogen column density is more difficult. It is
based on an empirically determined conversion factor $X_{\rm CO}$ applied to the
velocity integrated CO line. For SAMS the low value found for the large 
scale foreground clouds (de Vries et al. \cite{devries:etal87}) had 
been used.

SPIRE and MIPS data now provide an independent means to estimate the
dust temperatures and masses of SAMSs. In this paper the values were
derived adopting a gas-to-dust mass ratio of 100, which is well within the
values found for our Galaxy (Sodroski et al. \cite{sodroski:etal97}). 
These new estimates
confirm the previous ones, namely that SAMSs have low total hydrogen
column densities and, following from that, low masses, comparable to
that of Jupiter. Their temperatures are between 13 and 17\,K. SAMSs
therefore form most likely the very low mass end of molecular clouds
in the Galactic cirrus. As such they can provide valuable clues for
the understanding of the formation or destruction of molecular clouds
in our Galaxy.

Note that this set of observations however does not provide
independent clues on the metallicty of the clouds for the following
reason: the $X_{\rm CO}$ factor is found to increase with decreasing
metallicity (e.g. Wilson \cite{wilson95}, Barone et
al. \cite{barone:etal00}). For a given CO intensity a lower
metallicity would therefore result in a higher H$_2$ column
density. Similarly, the gas-to-dust ratio increases with decreasing
metallicity. For a lower metallicity a given infrared intensity would
thus result in a higher H$_2$ column density, too.

\subsection{Tidal arms or Galactic cirrus clouds?}

The dust temperatures and column densities of the clouds examined in
this paper are very similar. Based on the assumption of a Galactic
dust-to-gas ratio total hydrogen column densities are found to be in
the range between 1.8 and $7.2\cdot 10^{20}$\,cm$^{-2}$. Dust
temperatures are between 13 to 17\,K. There is no obvious difference
between the individual clouds. So there is no way to distinguish
between Galactic and extragalactic origin based on dust color ratios
or infrared intensity alone.

In the region towards M\,81 at least three out of six clouds are
clearly detected in CO lines of Galactic origin. The other
filamentary clouds have similar structures and similar intensities in
the SPIRE bands. They are thus candidates for small-area molecular
structures which have however to be verified by deep CO observations.

Comparison of deep optical images (e.g. given in Sollima et
al. \cite{sollima:etal10}) with the SPIRE 250\,$\mu$m map shows that
the clouds detected in CO (SAMS2, 4, and 6) are morphological very
similar. These dust clouds form part of the structure known as Arp's
ring or loop (Arp \cite{arp65}). The narrow CO linewidth and the close
agreement of the line of sight velocities with local HI gas
(Heithausen \cite{heithausen02}, \cite{heithausen04}) clearly rule out
an extragalactic origin of this structure. We are thus just faced with
a chance superposition of foreground filaments forming a loop like
structure with the distant galaxy M\,81. The observations presented in
this paper thus confirm the conclusions of Sollima et
al. (\cite{sollima:etal10}) who, based on HI and dust observations,
found that the bulk of dust emission outside the main body of M\,81
is associated with Galactic cirrus clouds.
Davies et al. (\cite{davies:etal10}) describe a break-down of the
  HI-infrared relation on scales below 1 arcmin. This could be caused by
  the fact that the clouds become at least partly molecular, as can be seen by
the detection of CO towards some of these clouds.

Towards NGC\,3077 the situation is more complex: here we are faced
with one molecular cloud of Galactic and another molecular complex of
extragalactic origin. Both show up in all SPIRE bands with similar
color ratios. Our incomplete CO map presented in this
paper shows that the Galactic cloud SAMS1 is probably more extended
than previously thought. This suggests that even more of the dust
seen towards NGC\,3077 is located in the Galactic foreground. 
Walter et al. (\cite{walter:etal11}) attributed all SPIRE 
emission to the tidal tail of NGC\,3077,
whereas it is shown here that some of the IR emission is of Galactic origin.

\section{Conclusions}

The data presented in this paper show that the mere coincidence of HI
gas and dust emission on the same line of sight does not proof that
the clouds are really physically associated, even if the intensity
maxima partly coincide. CO gas is a much better indicator for an
association. The velocity information of the CO spectral lines
provides valuable clues on the origin of the gas and the associated
dust, either Galactic or extragalactic. To confirm or exclude either origin,
CO spectra should therefore cover a 
bandwidth which includes both Galactic and extragalactic velocities. 
Because in some cases such a configuration is not possible with
the receiver, the observer should consider to take at least some 
control spectra covering Galactic velocities.

Clearly, more extensive CO observations of the Galactic cirrus clouds
are demanded to study the importance of small area molecular structures.
Currently most of such surveys for molecular cirrus
clouds are done with moderate to low angular resolution (Heithausen et
al. \cite{heithausen:etal93}; Magnani et al. \cite{magnani:etal00};
Onishi et al. \cite{onishi:etal01}). Due to the small size of SAMSs
these observations should be conducted at high angular resolution and
high sensitivity. As shown in this paper emission at 250 to 500$\mu$m 
could be a good guide to find such clouds.

Finally, it would be interesting to study if there are more examples of
  dust clouds where their origin, either Galactic or extragalactic, is not
  clear.  Two such cases have been reported by Dirsch et
  al. (\cite{dirsch:etal03}; \cite{dirsch:etal05}) and by Cortese et
  al. (\cite{cortese:etal10}). Dirsch et al. found a single dense cloud of
  only 4$''$ diameter projected on a spiral arm of the galaxy NGC\,3269. Based
  on a study of the redding law of that object the authors conclude that it
  has a Galactic origin.  Cortese et al. describe a system of interacting
  galaxies similar to that of M\,81. Towards NGC\,4435/4438 they found a HI,
  CO, and dust cloud which could be interpreted as tidal stream or as Galactic
  foreground cirrus. Based on velocity information of HI and CO spectra the
  authors favour a Galactic origin. The width of their CO line of only $\Delta
  v=1.5$\,km\,s$^{-1}$ is similar to that of SAMS (Heithausen
  \cite{heithausen06}), i.e. more consistent with Galactic cirrus clouds than
  with extragalactic molecular complexes. It is certainly worthwhile to map
  the total extend of the molecular cloud in the NGC\,4435/4438 field and to
  search CO towards NGC\,3269, because they could be similar to the SAMSs
  described in this paper.

\begin{acknowledgements}
I thank Fabian Walter for critical comments on the manuscript.

SPIRE has been developed by a consortium of institutes led by Cardiff
University (UK) and including Univ. Lethbridge (Canada); NAOC (China);
CEA, LAM (France); IFSI, Univ. Padua (Italy); IAC (Spain); Stockholm
Observatory (Sweden); Imperial College London, RAL, UCL-MSSL, UKATC,
Univ. Sussex (UK); and Caltech, JPL, NHSC, Univ. Colorado (USA). This
development has been supported by national funding agencies: CSA
(Canada); NAOC (China); CEA, CNES, CNRS (France); ASI (Italy); MCINN
(Spain); SNSB (Sweden); STFC (UK); and NASA (USA).
\end{acknowledgements}

{}

\listofobjects

\end{document}